\documentclass[12pt]{article}
\usepackage{graphics}
\usepackage{epsfig,amsmath}
\input epsf
\begin{document}
\thispagestyle{empty}
\begin{center}

{\Large\bf{Strangeness asymmetry of the
nucleon \vspace*{2ex} \\ in the statistical parton model}} 

\vskip1.4cm
{\bf Claude Bourrely and Jacques Soffer}  
\vskip 0.3cm
Centre de Physique Th\'eorique, UMR 6207 \footnote{
Unit\'e Mixte de Recherche du CNRS et des Universit\'es
Aix-Marseille I et Aix-Marseille II et de l' Universit\'e du Sud
Toulon-Var, Laboratoire affili\'e \`a la FRUMAM.}
\\
CNRS-Luminy, Case 907\\
F-13288 Marseille Cedex 9 - France \\ 
\vskip 0.5cm
{\bf Franco Buccella}
\vskip 0.3cm
Dipartimento di Scienze Fisiche, Universit\`a di Napoli,\\
Via Cintia, I-80126, Napoli
and INFN, Sezione di Napoli, Italy
\vskip 2cm
{\bf Abstract}\end{center}
We extend to the strange quarks and antiquarks, the statistical approach of 
parton distributions and we calculate the strange quark asymmetry $s-\bar s$.
We find that the asymmetry is small, positive in the low $x$ region and 
negative in the high $x$ region.
In this framework, the polarized strange quarks and 
antiquarks distributions, which are obtained simultaneously, are found to be 
both negative for all $x$ values.

\vskip 1cm 

\vskip 1cm
\noindent CPT-2006/P.069

\noindent UNIV. NAPLES DSF 028/2006
\newpage
\section{Introduction}

Although strange quarks and antiquarks $s$ and $\bar s$ play a fundamental
role in the nucleon structure, they are much less known than the parton 
distribution functions (PDF) for the light quarks $u$ and $d$.
The measurements of the structure functions in deep inelastic scattering
(DIS) of charged leptons on hadrons provide the best informations on
$u$, $d$, whereas neutrino DIS and lepton-pair production
in hadron collisions put some constraints on the sea quark distributions
$\bar u$ and $\bar d$. Concerning the strange quarks, due to the fact that
the structure functions are largely dominated by $u$ and $d$, the extraction of
the small components $s$ and $\bar s$  is rather difficult. Therefore most of
the phenomelogical models for the PDF studies
use the simplifying assumption $s(x) = \bar s(x) = \kappa(\bar u + \bar d)/2$
(with $\kappa \sim 0.5$). However, nothing prevents $s(x) \ne \bar s(x)$ and
we will see how to achieve this inequality in the statistical parton model 
\cite{bbs1, bbs2, bbs3, bbs6}.

An experiment on neutrino (antineutrino) -nucleon charged-current DIS 
by the CCFR collaboration \cite{barzako95} at the Fermilab Tevatron
has measured the production of dimuon final states coming from a charm
quark fragmentation. This process involves the interaction of a neutrino
(antineutrino) with an $s$ ($\bar s$) or $d$ ($\bar d$) quarks, via a 
$W^{\pm}$ exchange,
which can be used to isolate their distributions. Since the contribution of 
the down to
charm production is Cabibbo suppressed, scattering off a strange quark is 
responsible for most of the total dimuon rate. Unfortunately, only
an average value of $s + \bar s$ was extracted from this experiment,
but the size of strange quark distribution was known for the first time. 
Later, the
NuTeV collaboration \cite{goncharov} has reached a greater accuracy by
a high-statistics dimuon measurement, allowing to study independent
information on $s$ and $\bar s$ and the difference $s - \bar s$.

On the theoretical side one of first attempt to separate the $s$ and
$\bar s$ distributions was investigated in a light-cone model
\cite{brodsky96} and more recently, other models based on nonperturbative
mechanisms were proposed \cite{cao03, ding04}. A global QCD fit to the CCFR 
and NuTeV dimuon data has shown a clear evidence that $s \ne \bar s$
\cite{olness}. In another approach based on
perturbative evolution in QCD at three loops
\cite{catani04}, one is able to generate a strange-antistrange asymmetry
although at the input scale $s = \bar s$.

In this Letter, we will show how to construct the strange and antistrange
quark distributions in the statistical parton model. Since according to our 
method, the basic distributions are the helicity dependent ones, 
$s^{\pm}$ and $\bar s^{\pm}$,
we will obtain simultaneously the unpolarized, 
$s=s^+ + s^-$, $\bar s = \bar s^+ + \bar s^-$, and the polarized PDF, 
$\Delta s = s^+ - s^-$, $\Delta \bar s = \bar s^+ - \bar s^-$. We will
also explain how to determine the few parameters involved. Our results will 
be compared with other theoretical predictions.
\section{Strange quark and antiquark distributions}

In the statistical approach the nucleon is viewed as a gas of massless partons 
(quarks, antiquarks, gluons) in equilibrium at a given temperature in a 
finite size volume. 
Like in our earlier work on the subject \cite{bbs1}, we propose to 
use a simple description of the parton distributions $p(x)$
proportional to $[\exp[(x - X_{0p})/{\bar x}] \pm 1]^{-1}$,
the {\it plus} sign for quarks and antiquarks, corresponds to a Fermi-Dirac
distribution and the {\it minus} sign for gluons, corresponds 
to a Bose-Einstein distribution.
Here $X_{0p}$ is a constant which plays the role of the {\it thermodynamical
potential} of the parton {\it p} and $\bar x$ 
is the {\it universal temperature}, which is the same for all partons.
Since quarks carry a spin-1/2, it is natural to consider that the basic
distributions are $q_i^{\pm}(x)$, corresponding to a quark of flavor 
{\it i} and helicity parallel or antiparallel to the nucleon helicity.
From the chiral structure of QCD, we have two important properties which
allow to relate quark and antiquark distributions and to restrict 
the gluon distribution:\\
- The potential of a quark $q_i^{h}$ of helicity {\it h} is opposite to the
potential of the corresponding antiquark $\bar q_i^{\,-h}$ of helicity 
{\it -h}, therefore $X_{0q}^h=-X_{0\bar q}^{-h}$.\\
- The potential of the gluon $G$ is zero $X_{0G}=0$.\\
The sum rules, coming from the quantum numbers of the proton, $u-\bar u =2$ 
and $d-\bar d =1$, give
rise to higher values for the potentials of the u's than for the d's. In fact 
we have found
$ X_{0u}^+ > X_{0d}^- \sim  X_{0u}^- > X_{0d}^+$, which is also consistent 
with the known facts that 
$\Delta u(x) > 0$ and  $\Delta d(x) < 0$.
This ordering leads immediately to some important
consequences for the light antiquarks, namely

i) $\bar d(x) > \bar u(x)$, the flavor symmetry breaking, which also follows
from the Pauli exclusion principle, whose effects are incorporated in the 
statistical model. 

ii) $\Delta \bar u(x) > 0$ and  $\Delta \bar d(x) < 0$. 

We now turn to the procedure to construct the strange quark distributions.
In the original version of the statistical parton model \cite{bbs1} we have
assumed that the unpolarized strange quark and antiquark distributions are 
equal and they can be described by a linear 
combination of light antiquark distributions at the input scale $Q^2_0$, namely
\begin{equation}
xs(x,Q_0^2)=x \bar s(x,Q_0^2)= \frac{1}{4}[x \bar u (x,Q_0^2) + x \bar
d(x,Q_0^2)]~,
\label{sqori}
\end{equation}
where the coefficient 1/4 is an average value of some current estimates. For
the corresponding polarized distributions a similar assumption was made, more
precisely
\begin{equation}
x\Delta s(x,Q_0^2)=x \Delta \bar s(x,Q_0^2)= \frac{1}{3}[x \Delta \bar d
(x,Q_0^2) -x \Delta \bar u(x,Q_0^2)]~,
\label{sbarqori}
\end{equation}
which leads to a large negative distribution, since $\Delta \bar d < 0$ and 
$\Delta \bar u > 0$ (See Fig. 18 of Ref. \cite{bbs1}).
In order to introduce a difference between $s$ and $\bar s$, here we follow
the procedure used earlier to built the light quarks PDF, with the recent 
improvement 
obtained from the extension to the transverse momentum of the PDF \cite{bbs6}.
So the strange quark distributions  $s^h(x,Q_0^2)$ of helicity $h =\pm$,
at the input energy scale $Q_0^2=4\mbox{GeV}^2$, have the following expressions
\begin{equation}
xs^{h}(x,Q_0^2)= \frac{A X_{0u}^{+} x^{b_ s}}{\exp[(x-X_{0s}^h)/{\bar x}]+1}
~\frac{\mbox{ln}\left(1 + 
\exp{[kX_{0s}^{h}/\bar x]}\right)}{\mbox{ln}\left(1 + 
\exp{[kX_{0u}^{+}/\bar x]}\right)}
+ \frac{{\tilde A_{s}} x^{\tilde b}}{\exp(x/{\bar x})+1}~,
\label{sq}
\end{equation}
and similarly \footnote{As mentioned above,
quarks and antiquarks are not independent due to the relation 
between the potentials $X_{0s}^h = -X_{0\bar s}^{-h}$.} for the antiquarks 
$\bar s^h(x,Q_0^2)$
\begin{equation}
x\bar s^{h}(x,Q_0^2)= \frac{\bar A (X_{0d}^{+})^{-1} x^{2b_ s}}
{\exp[(x+X_{0s}^{-h})/{\bar x}]+1}~\frac{\mbox{ln}\left(1 + 
\exp{[-kX_{0s}^{-h}/\bar x]}\right)}{\mbox{ln}\left(1 + 
\exp{[-kX_{0d}^{+}/\bar x]}\right)}
+ \frac{{\tilde A_{s}} x^{\tilde b}}{\exp(x/{\bar x})+1}~.
\label{sqbar}
\end{equation}
The value of the input energy scale is arbitrary and should not affect
the results which satisfy the $Q^2$ QCD evolution. Our choice was dictated
in Ref.~\cite{bbs1} by the existence of many accurate data at 
$Q^2=4\mbox{GeV}^2$.
The first term in the right hand side corresponds to the non diffractive
part, while the second is associated with a diffractive component common
to all distributions. The ratio of the logarithms originates simply from
our extension of the statistical distributions to the transverse degree
of freedom and justifies the presence of a multiplicative factor in the
Fermi-Dirac functions, first introduced in Ref.\cite{bbs1}, as was explained 
in Ref.\cite{bbs6}.
The above expressions involve some parameters already
determined in our previous works \cite{bbs1,bbs6}, which we recall now
\begin{eqnarray}
&& A = 1.74938,~\bar{A} = 1.90801, ~X^+_{0u}=0.46128, ~X^+_{0d}=0.22775, 
\nonumber \\
&& \bar{x}=0.09907, ~ \tilde{b}=-0.25347, ~ k=1.42
\,.
\label{valparam}
\end{eqnarray}

Therefore it remains only four free parameters to determine, namely the two 
potentials
$X^{\pm}_{0s}$, $b_s$ and the normalization of the diffractive part 
$\tilde A_{s}$.

In order to obtain these free parameters, we will use some constraints:
first, the nucleon does not have any strangeness quantum number, as a
consequence the asymmetry $[a^-]$ has to vanish for all $x$ values
\begin{equation}
[a^-] = \int_0^1[s(x) - \bar s(x)]dx = 0~,
\label{am}
\end{equation}
second, from the second Bjorken sum rule, the first moments of the polarized 
quark distributions must satisfy the relation
\begin{equation}
\Delta q_8 = \Delta u + \Delta \bar u + \Delta d + \Delta \bar d - 
2(\Delta s + \Delta \bar s) = 3F - D~,
\label{dq8}
\end{equation}
where $F$ and $D$ are the hyperon beta decay constants, so that 
$3F-D=0.579 \pm 0.008$.
From the values of the first moments of the light quarks calculated
in Ref. \cite{bbs1}, where quarks and antiquarks are related through their
potentials, we can deduce another constraint, namely
\begin{equation}
[a^+] = \int_0^1 [\Delta s(x) + \Delta \bar s(x)]dx = -0.04675~.
\label{ap}
\end{equation}
From the above discussion on the light quarks, it is now clear that the sum 
rule Eq.~(\ref{am})
will lead to strange potentials $X_{0s}^{\pm}$ smaller than $X_{0u}^{\pm}$ 
and $X_{0d}^{\pm}$.
This obvious expectation has been observed in several earlier works on the 
same subject, see for example Ref.~\cite{Bal}.
Similarly in order to satisfy Eq.(8), we anticipate that we
will find $X_{0s}^- > X_{0s}^+$.
To determine the free parameters, in addition to the above constraints, we
will use some experimental results 
obtained by the CCFR and NuTeV collaborations \cite{barzako95, goncharov} on 
the production of dimuons 
from neutrino and antineutrino scattering on iron.
We have performed a next-to-leading order (NLO) QCD analysis of the data, 
keeping the light quark distributions as in Ref.~\cite{bbs1}.
We have obtained the following values for the parameters,
$X_{0s}^+ = 0.08101$,  $X_{0s}^- = 0.20290$, $b_s = 2.05305$
and $\tilde A_{s} = 0.05762 $, all with an error of the order of few percents.
We observe that the chemical potentials
for the strange quarks are smaller than the potentials for light quarks
$u$ and $d$ and that $X_{0s}^-> X_{0s}^+$, like in the case of the $d$ quark. 
Due to the large value
of $b_s$, the contribution of the non diffractive component is strongly 
suppressed, in the small $x$ region.\\
Our results are displayed in Figs. 1-5 and the fit is rather satisfactory
since, as an indication, we have a $\chi^2/\mbox{dof}$ of the order of 1.5, 
compared to 2.75 if one uses instead,
the simplifying assumption Eq.~(1). We have also checked that in this ealier 
version, it is not possible
to reproduce the rapid rise of $s + \bar s$ at low $x$ and 
$Q^2=4\mbox{GeV}^2$ of the data, as shown in Fig.~1.

At the input scale $Q^2_0 = 4\mbox{GeV}^2$, $[a^-]=0$ is satisfied to a great 
accuracy 
and we have checked that this constraint is not affected by the $Q^2$ 
evolution. We also find
$[a^+] = -0.0221$, which is compatible with a recent HERMES determination 
\cite{jack06}, for this first moment in the measured region, $x>0.02$ and 
$<Q^2>=2.5 \mbox{GeV}^2$, namely 
$0.006 \pm 0.029(\mbox{stat}.) \pm 0.007(\mbox{sys}.)$.
 
For the first moment of the asymmetry
\begin{equation}
[S^-] = \int_0^1[s(x) - \bar s(x)]x dx~,
\label{asm}
\end{equation}
we have $[S^-] = -0.00194$, to be compared with the value
 $-0.0011\pm 0.0014$ found by the
NuTeV collaboration \cite{mason}, and with the allowed range extracted from a 
global QCD fit by CTEQ \cite{olness}
$-0.001 < [S^-] < 0.004$. The calculations in the light-cone
meson-baryon model, lead to two positive results, namely 
$0.0042 < [S^-] < 0.0106$ for the choice
of a Gaussian wave function or $0.0035 < [S^-] < 0.0087$ for a power-law wave 
function \cite{ding04}.

We now turn to discuss the main features of the distributions obtained from 
this fit and compare them with other theoretical models.
We show the unpolarized and polarized strange
quark distributions at the input scale $Q^2_0 = 4\mbox{GeV}^2$ in Fig.~6. 
We observe that
the distributions $s(x)$ and $\bar s(x)$ are almost identical for 
$x < 0.05$, because the diffractive component dominates largely, whereas 
$ s(x)$ is a little larger than $\bar s(x)$ for 
$0.05 < x < 0.25$ and $ s(x)< \bar s(x)$ for $0.25 < x < 1$. 
These features remain unchanged
for higher $Q^2$ values, as shown in Fig.~7 for the difference $s - \bar s$
plotted as a function of $x$, for  $Q^2 = 4, 20, 100\mbox{GeV}^2$. 
This pattern is similar to that one gets in the meson cloud model 
\cite{cao03} and also 
in another approach based on perturbative evolution in QCD at three loops
\cite{catani04}, although, in this later case the sign change occurs at a 
much smaller value of $x$.
On the contrary, in Ref.\cite{ding04} they found that $s(x) < \bar s(x)$ in 
the small $x$ region and $s(x) > \bar s(x)$ in the large $x$ region. 

Finally, both $\Delta s(x)$ and $\Delta \bar s(x)$ are negative for
all $x$, as shown in Fig.~6, in reasonable agreement with the results of 
Ref.~\cite{Bal}. 
This contradicts the expectation of the meson cloud model \cite{cao03}, so
it is clear that we need for a better measurement of the strange quark 
contribution to the nucleon spin, has was also stressed in Ref.~\cite{pate}.

\section{Conclusion}

We have investigated the possibility to introduce an asymmetry for the
strange quark distributions in the framework of a statistical parton
model. In the absence of direct precise experimental data, we have imposed
different unpolarized and polarized constraints on the distributions and an
extensive use of the recent results from CCFR and NuTeV.
The main results are that $s(x)-\bar s(x)$ is indeed small, as expected, 
positive
in the low $x$ region and negative for $x> 0.25$. Our approach has the unique 
feature to provide simultaneously the polarized distributions for strange
quarks and antiquarks which are found to be both negative for all $x$.
New results on the strange quarks distributions are welcome, because they
will produce further tests on the present determination and hopefully some 
improvement on them.


\begin{figure}[htb]
\begin{center}
\leavevmode {\epsfysize=14.cm \epsffile{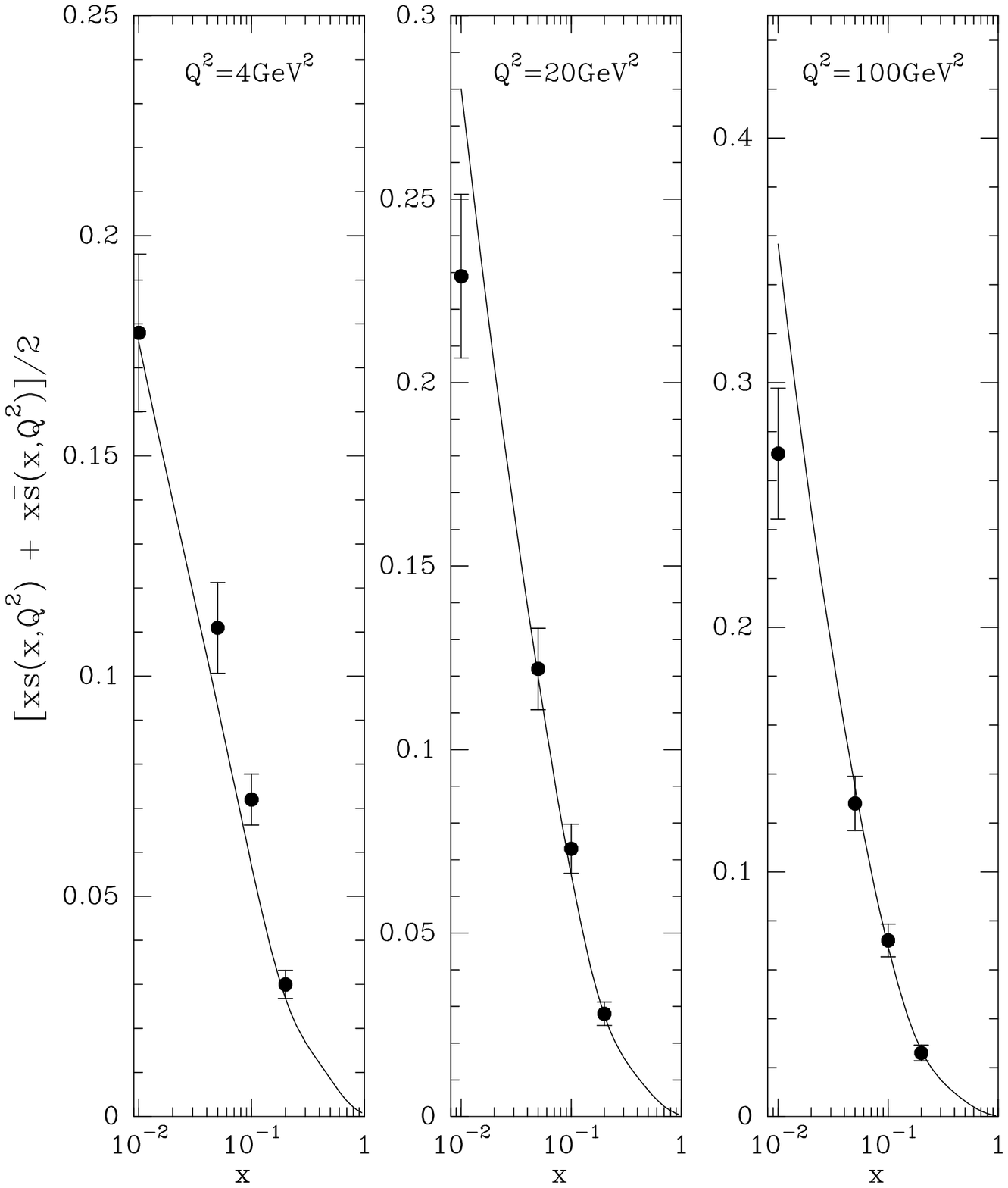}}
\end{center}
  \vspace*{-5mm}
\caption[*]{\baselineskip 1pt
The average strange quark antiquark distributions determined at NLO as 
a function of $x$ for  $Q^2 = 4, 20, 100\mbox{GeV}^2$. Data from CCFR 
Collaboration \cite{barzako95}.}
\label{fi:average}
\vspace*{-1.0ex}
\end{figure}

\begin{figure}[htb]
\begin{center}
\leavevmode {\epsfysize=16.cm \epsffile{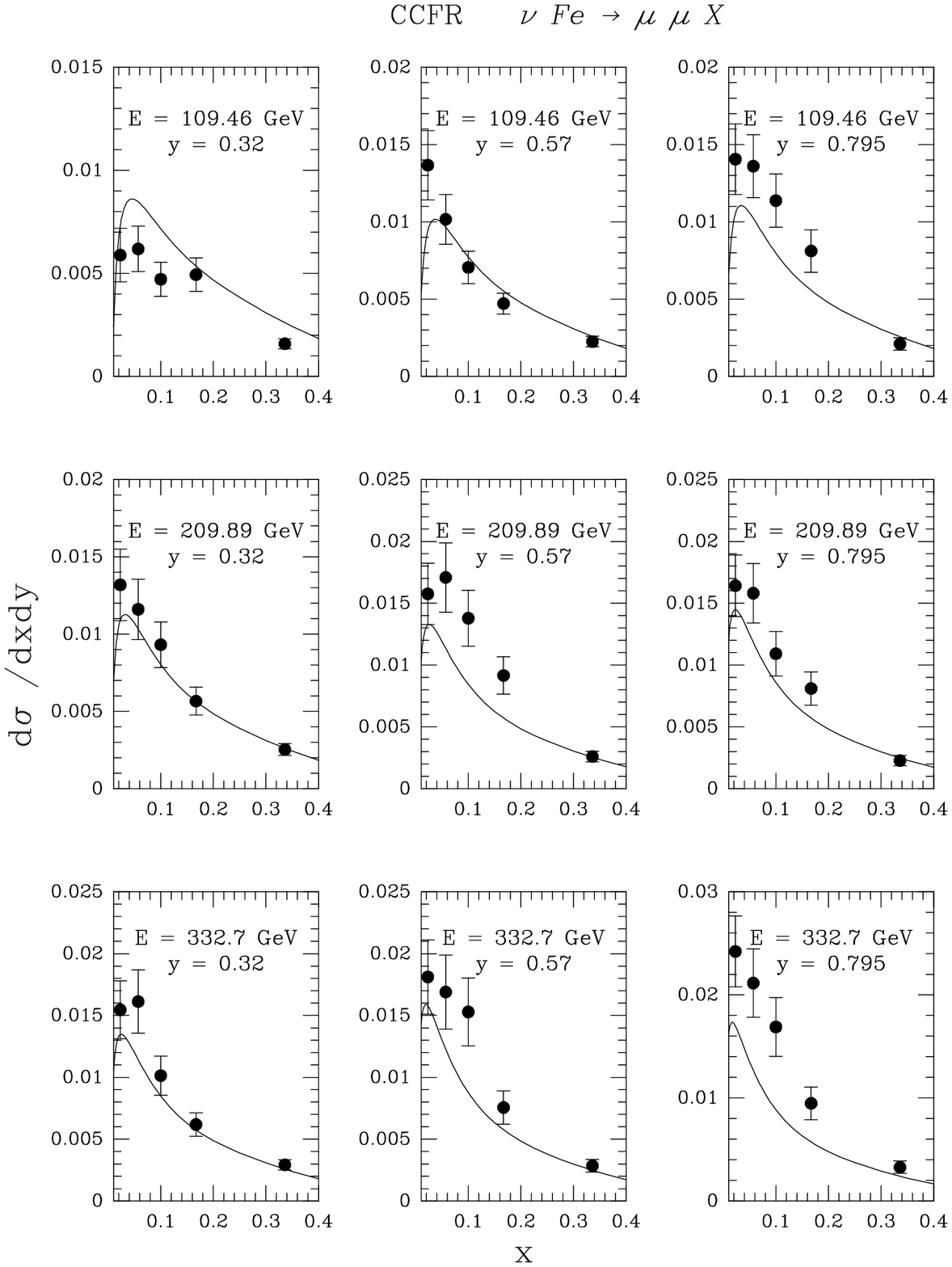}}
\end{center}
  \vspace*{-5mm}
\caption[*]{\baselineskip 1pt
Comparison of the CCFR $\nu$ data \cite{goncharov} to the result of the fit 
for $d\sigma/dxdy$, 
in units of charged-current $\sigma$, for various kinematic ranges in energy, 
$x$ and $y$.}
\label{fi:ccfrnu}
\vspace*{-1.0ex}
\end{figure}

\begin{figure}[htb]
\begin{center}
\leavevmode {\epsfysize=16.cm \epsffile{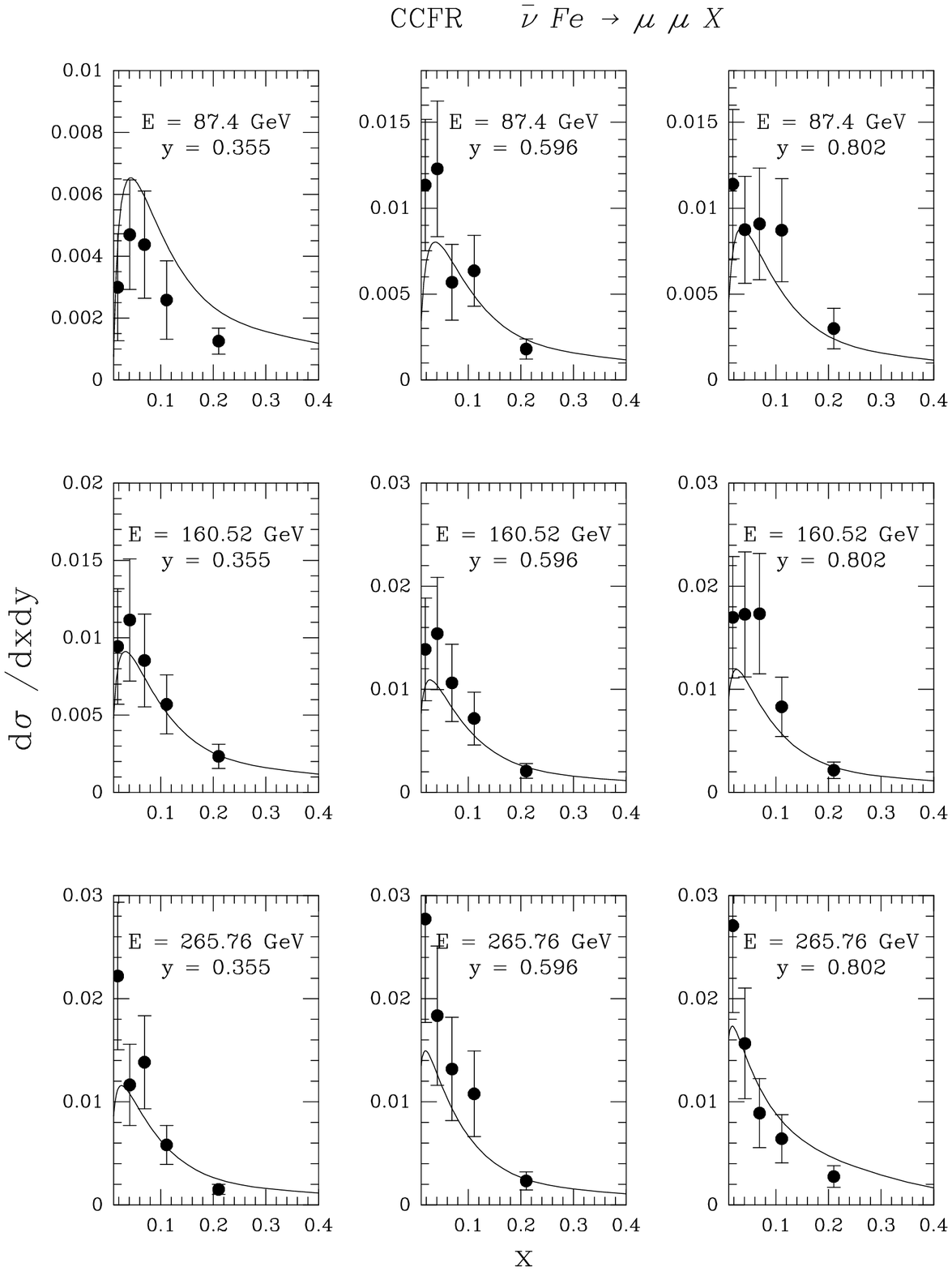}}
\end{center}
  \vspace*{-5mm}
\caption[*]{\baselineskip 1pt
Comparison of the CCFR $\bar \nu$ data \cite{goncharov} to the result of the 
fit for $d\sigma/dxdy$ 
in units of charged-current $\sigma$, for various kinematic ranges in energy, 
$x$ and $y$.}
\label{fi:ccfranu}
\vspace*{-1.0ex}
\end{figure}

\begin{figure}[htb]
\begin{center}
\leavevmode {\epsfysize=16.cm \epsffile{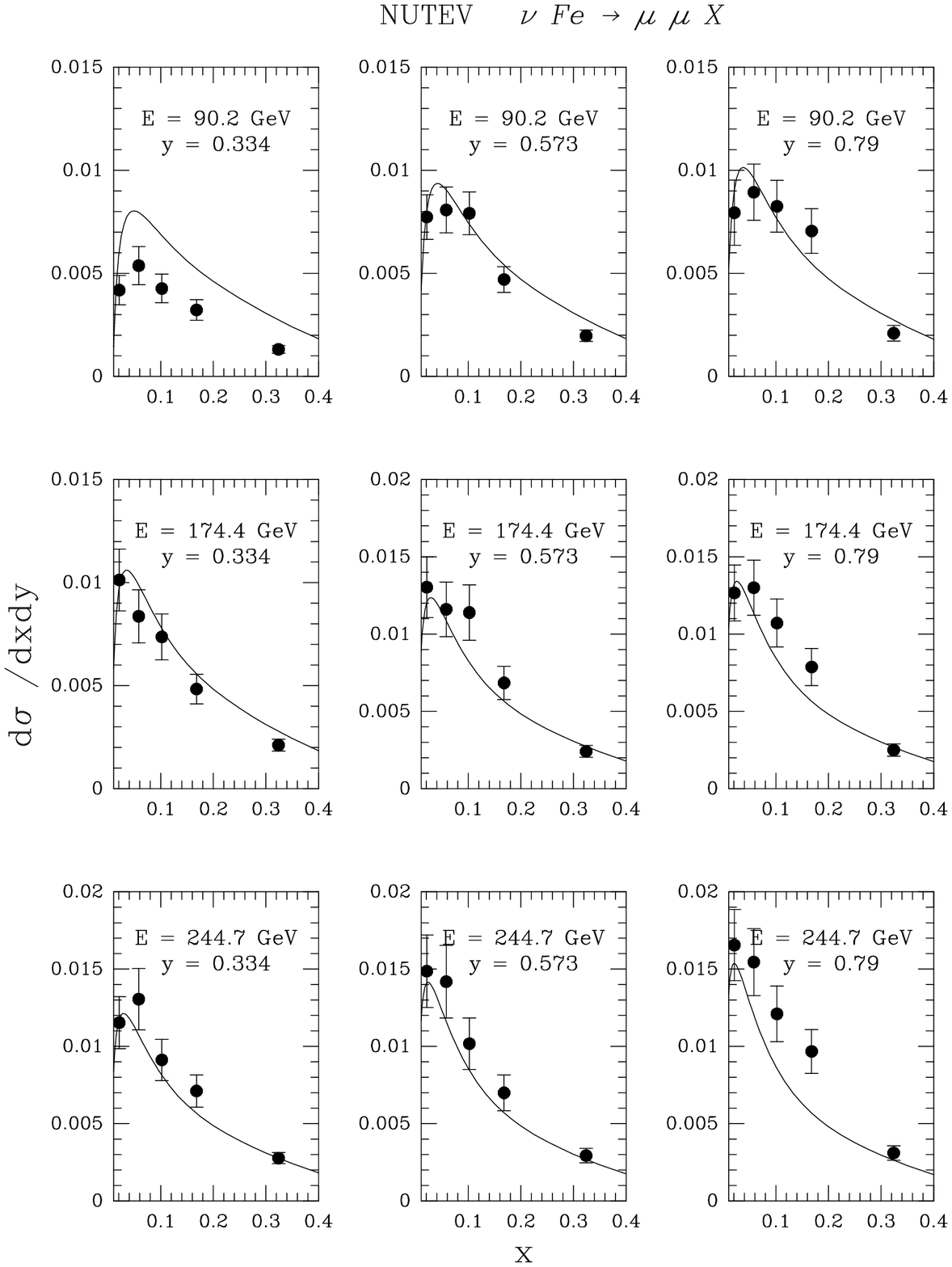}}
\end{center}
  \vspace*{-5mm}
\caption[*]{\baselineskip 1pt
Comparison of the NuTeV $\nu$ data \cite{goncharov} to the result of the fit 
for $d\sigma/dxdy$,
in units of charged-current $\sigma$, for various kinematic ranges in energy, 
$x$ and $y$.}
\label{fi:nutevnu}
\vspace*{-1.0ex}
\end{figure}

\begin{figure}[htb]
\begin{center}
\leavevmode {\epsfysize=16.cm \epsffile{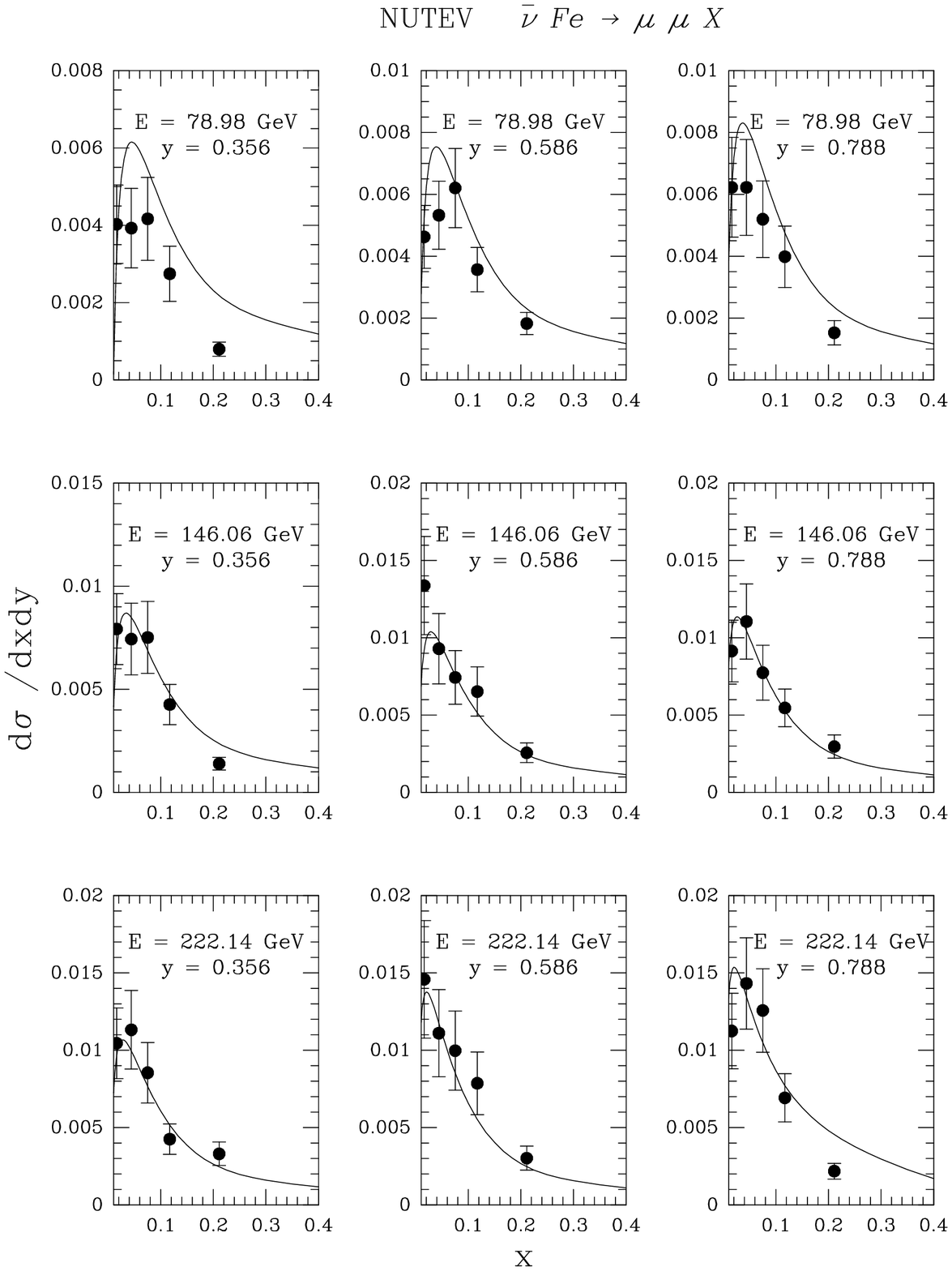}}
\end{center}
  \vspace*{-5mm}
\caption[*]{\baselineskip 1pt
Comparison of the NuTeV $\bar \nu$ data \cite{goncharov} to the result of the 
fit for $d\sigma/dxdy$,
in units of charged-current $\sigma$, for various kinematic ranges in energy, 
$x$ and $y$.}
\label{fi:nutevanu}
\vspace*{-1.0ex}
\end{figure}

\begin{figure}[htb]
\begin{center}
\leavevmode {\epsfysize=14.cm \epsffile{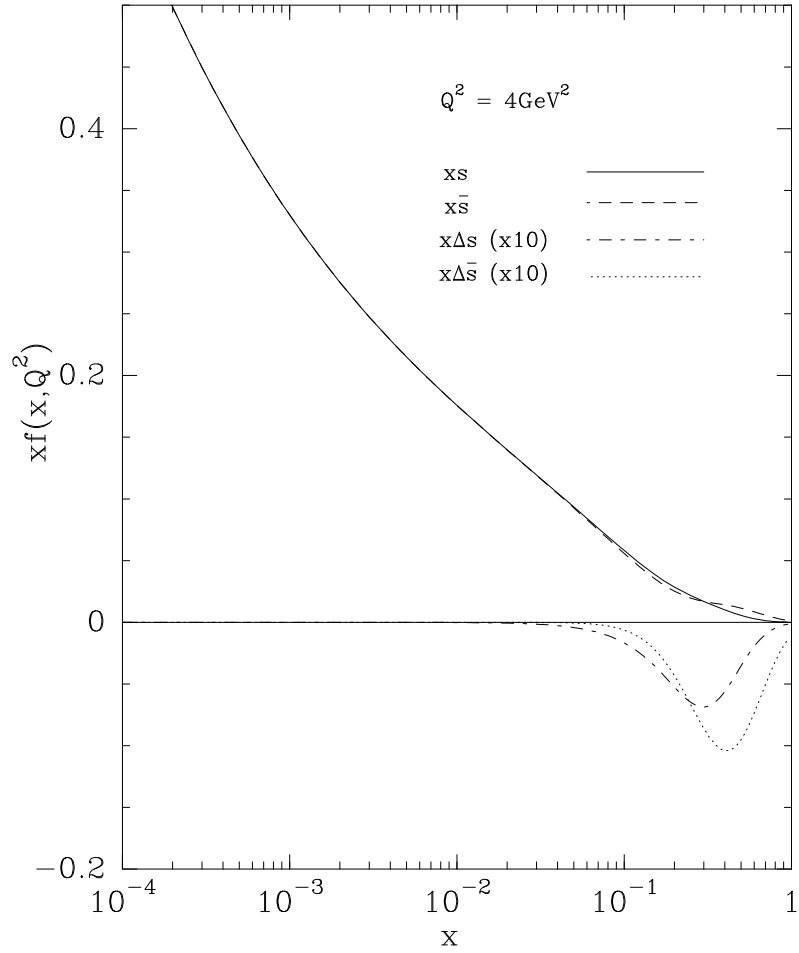}}
\end{center}
  \vspace*{-5mm}
\caption[*]{\baselineskip 1pt
The unpolarized and polarized strange quark and antiquark distributions 
determined at NLO as a function of $x$ for  $Q^2 = 4\mbox{GeV}^2$.}
\label{fi:strang}
\vspace*{-1.0ex}
\end{figure}

\begin{figure}[htb]
\begin{center}
\leavevmode {\epsfysize=14.cm \epsffile{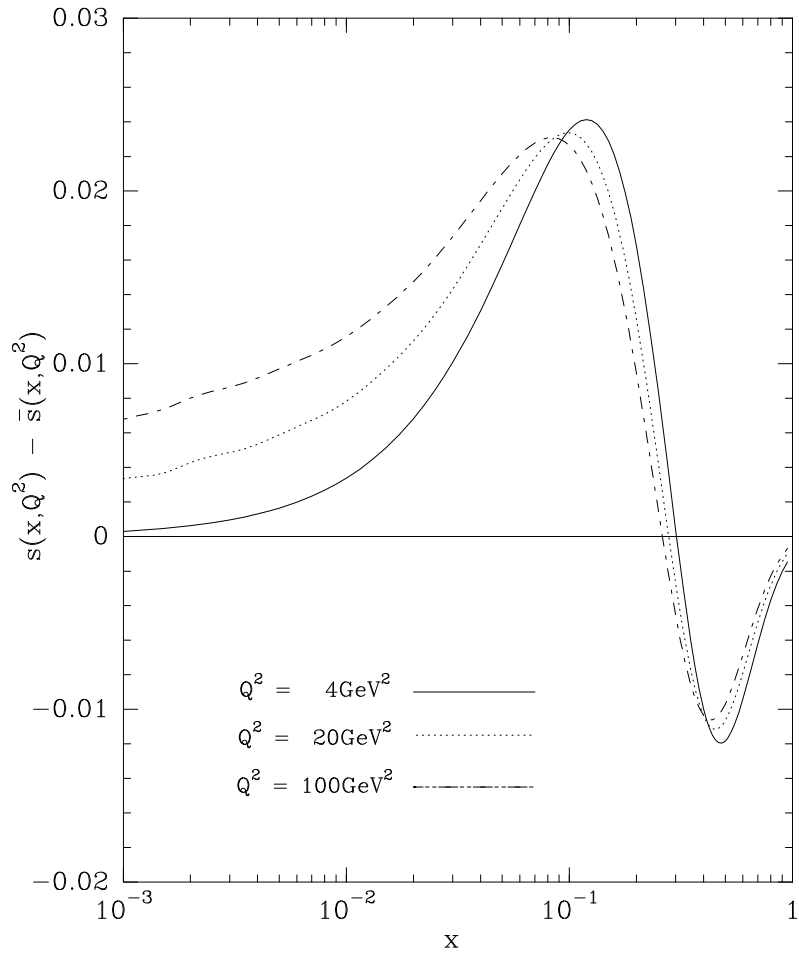}}
\end{center}
  \vspace*{-5mm}
\caption[*]{\baselineskip 1pt
The difference $s - \bar s$ quark distributions  determined at NLO as 
a function of $x$ for  $Q^2 = 4, 20, 100\mbox{GeV}^2$.}
\label{fi:diffs}
\vspace*{-1.0ex}
\end{figure}

\end{document}